\documentclass[12pt]{article}
\usepackage{amssymb}
\usepackage{latexsym,amsfonts,amsmath,amssymb,wasysym,enumerate,epsfig}
\usepackage{mathrsfs}
\renewcommand{\theequation}{\arabic{section}.\arabic{equation}}

\begin{document}



\def\a{\alpha}
\def\b{\beta}
\def\d{\delta}
\def\e{\epsilon}
\def\g{\gamma}
\def\h{\mathfrak{h}}
\def\k{\kappa}
\def\l{\lambda}
\def\o{\omega}
\def\p{\wp}
\def\r{\rho}
\def\t{\tau}
\def\s{\sigma}
\def\z{\zeta}
\def\x{\xi}
\def\V={{{\bf\rm{V}}}}
 \def\A{{\cal{A}}}
 \def\B{{\cal{B}}}
 \def\C{{\cal{C}}}
 \def\D{{\cal{D}}}
\def\G{\Gamma}
\def\K{{\cal{K}}}
\def\O{\Omega}
\def\R{\bar{R}}
\def\T{{\cal{T}}}
\def\L{\Lambda}
\def\f{E_{\tau,\eta}(sl_2)}
\def\E{E_{\tau,\eta}(sl_n)}
\def\Zb{\mathbb{Z}}
\def\Cb{\mathbb{C}}

\def\R{\overline{R}}

\def\beq{\begin{equation}}
\def\eeq{\end{equation}}
\def\bea{\begin{eqnarray}}
\def\eea{\end{eqnarray}}
\def\ba{\begin{array}}
\def\ea{\end{array}}
\def\no{\nonumber}
\def\le{\langle}
\def\re{\rangle}
\def\lt{\left}
\def\rt{\right}

\newtheorem{Theorem}{Theorem}
\newtheorem{Definition}{Definition}
\newtheorem{Proposition}{Proposition}
\newtheorem{Lemma}{Lemma}
\newtheorem{Corollary}{Corollary}
\newcommand{\proof}[1]{{\bf Proof. }
        #1\begin{flushright}$\Box$\end{flushright}}

\baselineskip=20pt

\newfont{\elevenmib}{cmmib10 scaled\magstep1}
\newcommand{\preprint}{
   \begin{flushleft}
   \end{flushleft}\vspace{-1.3cm}
   \begin{flushright}\normalsize
   \end{flushright}}
\newcommand{\Title}[1]{{\baselineskip=26pt
   \begin{center} \Large \bf #1 \\ \ \\ \end{center}}}
\newcommand{\Author}{\begin{center}
   \large \bf
Xiaotian Xu${}^{a,b}$, Kun Hao${}^{a,b}$,~~Tao Yang${}^{a,b}$\footnote{Corresponding author:
yangt@nwu.edu.cn},~~Junpeng Cao${}^{c,d,e}$,~~Wen-Li Yang${}^{a,b,f}\footnote{Corresponding author:
wlyang@nwu.edu.cn}$ and Kangjie Shi${}^{a,b}$
 \end{center}}
\newcommand{\Address}{\begin{center}

     ${}^a$Institute of Modern Physics, Northwest University,
     Xian 710069, China\\
     ${}^b$Shaanxi Key Laboratory for Theoretical Physics Frontiers,  Xian 710069, China\\
     ${}^c$Beijing National Laboratory for Condensed Matter
           Physics, Institute of Physics, Chinese Academy of Sciences, Beijing
           100190, China\\
     ${}^d$Collaborative Innovation Center of Quantum Matter, Beijing,
     China\\
     ${}^e$School of Physical Sciences, University of Chinese Academy of
Sciences, Beijing, China\\
     ${}^f$Beijing Center for Mathematics and Information Interdisciplinary Sciences, Beijing, 100048,  China

   \end{center}}
\newcommand{\Accepted}[1]{\begin{center}
   {\large \sf #1}\\ \vspace{1mm}{\small \sf Accepted for Publication}
   \end{center}}

\preprint
\thispagestyle{empty}
\bigskip\bigskip\bigskip

\Title{Bethe ansatz solutions of the $\tau_2$-model
 with arbitrary boundary fields } \Author

\Address
\vspace{1cm}

\begin{abstract}
The quantum $\tau_2$-model with generic site-dependent inhomogeneity  and arbitrary  boundary fields is studied via
the off-diagonal Bethe Ansatz method. The eigenvalues of the corresponding transfer matrix are given in terms of
an inhomogeneous $T-Q$ relation, which is  based on the operator product identities among the fused transfer matrices and the asymptotic behavior of the transfer matrices. Moreover, the associated Bethe Ansatz equations are also obtained.

\vspace{1truecm} \noindent {\it PACS:} 75.10.Pq, 03.65.Vf, 71.10.Pm

\noindent {\it Keywords}: Integrable system; reflection equation; Bethe
Ansatz; $T-Q$ relation
\end{abstract}
\newpage
\section{Introduction}
\label{intro} \setcounter{equation}{0}
The finite-size inhomogeneous $\tau_2$-model also known as the Baxter-Bazhanov-Stroganov model (BBS model) \cite{Baxter04, Baxter89, Bazhanov90, Baxter90} is a $N$-state spin lattice model, which is intimately related to some other integrable models under certain parameter constraints such as the chiral Potts model \cite{Gehlen85, Alcaraz86, Bashilov80, Au87, McCoy87, Perk88} and the relativistic quantum Toda chain model \cite{Rui90}. Lots of papers have appeared to explain such connections and many efforts have been made to calculate the eigenvalues of the chiral Potts model by solving the $\tau_2$-model with a recursive functional relation \cite{Baxter90, Albertini89, RJ05}. The $\tau_2$-model is a simple quantum integrable models associated with cyclic representation of the Wely algebra. Although its integrability has been proven \cite{Bazhanov90} for decades, there is still no effective method to solve the model completely due to lack of a simple $Q$-operator solution in terms of Baxter's $T-Q$ relation. In fact, the $Q$-operator is a very complicated function defined in high genus space and its concrete expression is hard to be derived. Very recently, Paul Fendley had found a ``parafermionic" way to diagonalise a simple solvable Hamiltonian associated with the chiral Potts model \cite{Fendley14}. Subsequently, this method was generalised  to solve the $\tau_2$-model with particular open boundaries \cite{Baxter2014, AuY14, AuY16}. Until very recently, the Bethe Ansatz solution of the periodic $\tau_2$-model with generic site-dependent inhomogeneity was obtained by constructing an inhomogeneous $T-Q$ relation with polynomial $Q$-functions (i.e. off-diagonal Bethe Ansatz method (ODBA) \cite{Cao1, yu15, Xu15}), which provides a perspective to investigate the $\tau_2$-model with generic open boundary condition. Moreover, such a solution allows the authors \cite{Zha16} further to retrieve the corresponding Bethe states.

The aim of this paper is to explicitly construct the eigenvalues of the transfer matrix for the open  $\tau_2$-model
with the most generic inhomogeneity, where we generalise the ODBA method to solve the open inhomogeneous $\tau_2$-model with arbitrary integrable open boundary condition in combination with the fusion technique. By introducing an off-diagonal term in the conventional $T-Q$ relation (i.e., the inhomogeneous $T-Q$ relation), we obtain the spectrum of the generic open $\tau_2$-model and the associated Bethe Ansatz equations.

The outline of this paper is as follows. In Section 2, we begin with a brief introduction of the  fundamental transfer matrix. In Section 3, we study the properties of the transfer matrix and employing the so-called fusion procedure \cite{Kulish1981, Kulish1982, Kirillov1982} to construct the higher-spin transfer matrices, which obey an infinite fusion hierarchy. In Section 4, we obtain the truncation identity for the fused transfer matrices when the bulk anisotropy value takes the special case $\eta=\frac{2i\pi}{p}$ and the exact functional relations of the fundamental transfer matrix. In Section 5, we give the eigenvalues of the transfer matrix in terms of
some inhomogeneous $T-Q$ relation and the associated Bethe Ansatz equation. In the last Section, we summarize our results and give some discussions.
Some detailed technical calculations are given  in appendices A and B.


\section{ Transfer matrix}
\label{XXZ} \setcounter{equation}{0}
\setcounter{equation}{0}

Let us fix an odd integer $p$ such that $p\geq 3$, and let ${\rm\bf V}$ be a $p$-dimensional vector space (i.e. the local
Hilbert space) with an orthonormal basis $\{|m\rangle
|m\in\mathbb{Z}_p\}$. Define two $p\times p$ matrices $X$ and $Z$
which act on the basis as
\bea
X|m\rangle=q^m|m\rangle,\quad Z|m\rangle=|m+1\rangle,\quad
m\in\mathbb{Z}_p,
\eea where $q\equiv e^{-\eta}$ is a $p$-root of unity (i.e., $q^p=1$). The embedding operators
$\{X_n,Z_n|n=1,\cdots,N\}$ denote the generators of the ultra-local
Weyl algebra:
\bea X_nZ_m=q^{\delta_{nm}}Z_mX_n,\quad
X_n^p=Z_n^p=1,\quad \forall n, m\in\{1, \cdots, N\}. \label{Weyl}
\eea
It has been shown that the $\tau_2$-model can be described by a quantum integrable spin chain \cite{Bazhanov90}.
In order to construct the monodromy matrix, one need to introduce the $L$-operators for each site of the quantum chain.
The associated $L$-operator $L_n(u)\in {\rm End}({\rm \bf C^2}\otimes{\rm\bf V})$ defined in
the most general cyclic representation of $U_q(sl_2)$,  is  given by \cite{Bazhanov90}
\bea
L_{n}(u)&=&\left(
\begin{array}{cc}
  e^{u}d^{(+)}_n X_n+e^{-u}d^{(-)}_nX^{-1}_n & (g^{(+)}_nX^{-1}_n+g^{(-)}_n X_n)Z_n \\[6pt]
  (h^{(+)}_nX^{-1}_n+h^{(-)}_n X_n)Z^{-1}_n & e^{u}f^{(+)}_nX^{-1}_n+e^{-u}f^{(-)}_n X_n
\end{array}\right)\no\\[6pt]
&=&\left(\begin{array}{cc}
                           A_n(u) & B_n(u) \\[6pt]
                           C_n(u) & D_n(u)
                         \end{array}
\right),\quad n=1,\ldots, N,\label{L-operator} \eea
 where
$d^{(+)}_n$, $d^{(-)}_n$, $g^{(+)}_n$, $g^{(-)}_n$, $h^{(+)}_n$,
$h^{(-)}_n$, $f^{(+)}_n$ and $f^{(-)}_n$ are some parameters
associated with each site. These parameters are subjected to two
constraints, \bea g^{(-)}_nh^{(-)}_n=f^{(-)}_nd^{(+)}_n,\quad
g^{(+)}_nh^{(+)}_n=f^{(+)}_nd^{(-)}_n,\quad
n=1,\cdots,N,\label{Constraint-1}
\eea
which ensure that the above L-operator $L_n$ satisfies  the Yang-Baxter algebra \cite{Bazhanov90},
\bea
R(u-v)(L_{n}(u)\otimes 1)(1\otimes L_{n}(v))=(1\otimes
L_{n}(v))(L_{n}(u)\otimes 1)R(u-v),\quad n=1,\ldots,N.
\label{Rll-relation}
\eea
The associated $R$-matrix $R(u)\in {\rm End}({\rm\bf C^2}\otimes {\rm\bf C^2})$ is
the well-known six-vertex $R$-matrix given by
\bea
R(u)=\left(
\begin{array}{cccc}
  \sinh(u+\eta) & 0 & 0 & 0 \\
  0 & \sinh u & \sinh \eta & 0 \\
  0 & \sinh \eta & \sinh u & 0 \\
  0 & 0 & 0 & \sinh(u+\eta)
\end{array}\right),\label{R-matrix}
\eea
with the crossing parameter $\eta$ taking the special values
\bea
\eta=2i\pi /p,\quad p=2l+1,\quad l=1,2,\cdots.\label{root}
\eea
The R-matrix satisfies the quantum Yang-Baxter equation (QYBE) \cite{Bax82,Kor93}
and becomes some projectors when the spectral parameter $u$ takes some special values as
\bea
&&\mbox{Antisymmetric-fusion conditions}:\,R(-\eta)=-2\sinh\eta P^{(-)},\label{Fusion-Con-1}\\[6pt]
&& \mbox{Symmetric-fusion conditions}:\, R(\eta) = 2\sinh\eta\,{\rm
Diag}(\cosh\eta,1,1,\cosh\eta)\,P^{(+)},\label{Fusion-Con-2} \eea
where $P^{(+)}$ ($P^{(-)}$) is the symmetric (anti-symmetric)
projector of the tensor space ${\rm\bf C^2}\otimes {\rm\bf C^2}$.
Associated with the local $L$-operators $\{L_n(u)|n=1,\ldots,N\}$ given by (\ref{L-operator}), let us introduce the one-row monodromy matrix $T(u)$
\bea T(u)= \left(\begin{array}{cc}
                           {\bf A}(u) & {\bf B}(u) \\[6pt]
                           {\bf C}(u) & {\bf D}(u)
                         \end{array}
\right)=L_N(u)\,L_{N-1}(u)\,\cdots L_1(u).\label{Mono-T}
\eea
The local relations (\ref{Rll-relation}) imply that the monodromy matrix $T(u)$ also satisfies the Yang-Baxter algebra
\bea
R(u-v)(T(u)\otimes 1)(1\otimes T(v))=(1\otimes T(v))(T(u)\otimes 1)R(u-v),\label{QYBE}
\eea
which ensures the integrability of the $\tau_2$-model with the periodic boundary condition \cite{Bazhanov90}.

Integrable open chain can be constructed as follows \cite{Skl1988}.
Let us introduce a pair of $K$-matrices $K^-(u)$ and $K^+(u)$. The
former satisfies the reflection equation (RE) \cite{Che84}
 \bea &&R_{12}(u_1-u_2)K^-_1(u_1)R_{21}(u_1+u_2)K^-_2(u_2)\no\\
 &&~~~~~~=
K^-_2(u_2)R_{12}(u_1+u_2)K^-_1(u_1)R_{21}(u_1-u_2),\label{RE-V}\eea
and the latter  satisfies the dual RE \bea
&&R_{12}(u_2-u_1)K^+_1(u_1)R_{21}(-u_1-u_2-2\eta)K^+_2(u_2)\no\\
&&~~~~~~=
K^+_2(u_2)R_{12}(-u_1-u_2-2\eta)K^+_1(u_1)R_{21}(u_2-u_1).
\label{DRE-V}\eea
In order to construct the associated open spin chain, let us introduce the $\hat{L}(u)$ in the form of
\bea
\hat{L}_{n}(u)&=&\left(
\begin{array}{cc}
  e^{-u-\eta}f^{(+)}_nX_n^{-1}+e^{u+\eta}f^{(-)}_nX_n & -(g^{(+)}_nX_n^{-1}+g^{(-)}_nX_n)Z_n \\
  -(h^{(+)}_nX_n^{-1}+h^{(-)}_nX_n)Z_n^{-1} &  e^{-u-\eta}d^{(+)}_nX_n+e^{u+\eta}d^{(-)}_nX_n^{-1}
\end{array}
\right)\no\\[6pt]
&=&\left(\begin{array}{cc}D_n(-u-\eta)&-B_n(-u-\eta)\\
-C_n(-u-\eta)&A_n(-u-\eta)\end{array}
\right).\label{L-2}
\eea
It is easy to check that $L_{n}(u)$ enjoys the crossing property
\bea
L_{n}(u)=\sigma^{y}\hat{L}_{n}^{t}(-u-\eta)\sigma^{y},\quad n=1,\ldots,N\label{crosing-unitarity},
\eea and the inverse relation
\bea
L_{n}(u)\,\hat{L}_{n}(-u)={\rm Det}_{q}\{L_{n}(u)\}\times {\rm id},\quad n=1,\ldots,N,\no
\eea
where the function ${\rm Det}_{q}\{L_{n}(u)\}$ is the quantum determinant (which will be given by below (\ref{det-t})).
Associated with the local $L$-operators $\{\hat{L}_n(u)|n=1,\ldots,N\}$ given by (\ref{L-2}), let us introduce
another one-row monodromy matrix $\hat{T}(u)$ (c.f., (\ref{Mono-T}))
\bea \hat{T}(u)= \left(\begin{array}{cc}
                           {\bf {\hat{A}}}(u) & {\bf {\hat{B}}}(u) \\[6pt]
                           {\bf {\hat{C}}}(u) & {\bf {\hat{D}}}(u)
                         \end{array}
\right)=\hat{L}_1(u)\,\hat{L}_{2}(u)\,\cdots \hat{L}_N(u).\label{Mono-{T1}}
\eea
For the system with integrable open boundaries, the transfer matrix $t(u)$ of the $\tau_2$-model with open boundaries can be constructed as \cite{Skl1988}
\bea
t(u)=tr\{K^+(u)T(u)K^{-}(u)\hat{T}(u)\},\label{transfer}
\eea
where $tr$ denotes trace over ``auxiliary space". The quadratic relation (\ref{QYBE}) and (dual) reflection equations (\ref{RE-V}) and (\ref{DRE-V})
lead to the fact that the transfer matrix $t(u)$ of the $\tau_2$-model with different spectral parameters are mutually commutative \cite{Skl1988},
i.e., $[t(u),t(v)]=0$, which ensures the integrability of the model by treating $t(u)$ as the generating functional of the conserved quantities.

In this paper, we consider  the most generic non-diagonal $K^-(u)$ matrix found in Refs. \cite{Vega94, Ghoshal94}, which is in the form of
\bea
K^-(u)&=&\lt(\begin{array}{ll}K_{11}^{-}(u)&K_{12}^{-}(u)\\[6pt]
K_{21}^{-}(u)&K_{22}^{-}(u)\end{array}\rt),
\eea
with
\bea
K_{11}^{-}(u)&=&2[\sinh(\alpha_-)\cosh(\beta_-)\cosh(u)+\cosh(\alpha_-)\sinh(\beta_-)\sinh(u)],\no\\[6pt]
K_{22}^{-}(u)&=&2[\sinh(\alpha_-)\cosh(\beta_-)\cosh(u)-\cosh(\alpha_-)\sinh(\beta_-)\sinh(u)],\no\\[6pt]
K_{12}^{-}(u)&=&e^{\theta_-}\sinh(2u),~~~~K_{21}^{-}(u)=e^{-\theta_-}\sinh(2u),
\label{K_-(u)}\eea
where $\alpha_-, \beta_-$, and $\theta_-$ are three free boundary parameters.
The most generic non-diagonal $K$-matrix $K^+(u)$ is given by
\bea
K^+(u)&=&K^{-}(-u-\eta)|_{(\alpha_-, \beta_-, \theta_-)\rightarrow(-\alpha_+, -\beta_+, \theta_+)}.\label{K +(u)}
\eea
We note that the two $K$-matrices possess the following properties
\bea
K^{\mp}(u+i\pi)=-\sigma^{z}K^{\mp}(u)\sigma^{z}\label{K-periodicity},
\eea
\bea
K^-(0)=\frac{1}{2}tr(K^-(0))\times {\rm id},\quad
K^-(\frac{i\pi}{2})=\frac{1}{2}tr(K^-(\frac{i\pi}{2})\s^z)\times
\s^z,\label{k-special value}
\eea
where $\sigma^{\alpha}$ with $\alpha=x,\,y,\,z$ are the Pauli matrices.


\section{ Properties of the transfer matrix}
\label{T-QR} \setcounter{equation}{0}
\subsection{Asymptotic behaviors and average values}

Based on the explicit expressions (\ref{L-operator}) and (\ref{L-2}) of the L-operators, the generic boundary matrices (\ref{K_-(u)})-(\ref{K +(u)}), and the definition (\ref{Mono-T})-(\ref{Mono-{T1}}) of the monodromy matrices, we note that the transfer matrix $t(u)$ given by (\ref{transfer})
has the asymptotic behavior,
\bea
\lim_{u\rightarrow \pm\infty}t(u)&=&-\frac{1}{4}e^{\pm \{(2N+4)u+(N+2)\eta\}}\lt\{e^{\theta_+-\theta_-}\,F^{(+)}F^{(-)}+e^{-\theta_++\theta_-}\,D^{(+)}D^{(-)}\rt\}
\times {\rm id},\label{Asymp}
\eea
where $D^{(\pm)}$ and $F^{(\pm)}$ are four constants related to the inhomogeneous parameters as follows,
\bea
D^{(\pm)}=\prod_{n=1}^Nd^{(\pm)}_n,\quad F^{(\pm)}=\prod_{n=1}^Nf^{(\pm)}_n.\label{Constant-1}
\eea
Moreover, we can calculate the special values of the associated transfer matrix at $u=0, \frac{i\pi}{2}$ with the help of the relations (\ref{k-special value}),
namely,
\bea
t(0)&=&-2^3\sinh\a_-\cosh\b_-\sinh\a_+\cosh\b_+\cosh\eta\,\no\\[6pt]
&&\times
\prod_{n=1}^N\Big(e^{-\eta}d_{n}^{(+)}f_{n}^{(+)}+e^{\eta}d_{n}^{(-)}f_{n}^{(-)}-e^{\eta}g_{n}^{(+)}h_{n}^{(-)}-e^{-\eta}g_{n}^{(-)}h_{n}^{(+)}\Big)\times
{\rm id},\\
t(\frac{i\pi}{2})&=&-2^3\cosh\a_-\sinh\b_-\cosh\a_+\sinh\b_+\cosh\eta\,\no\\[6pt]
&&\times
\prod_{l=1}^N\Big(e^{-\eta}d_{n}^{(+)}f_{n}^{(+)}+e^{\eta}d_{n}^{(-)}f_{n}^{(-)}+e^{\eta}g_{n}^{(+)}h_{n}^{(-)}+e^{-\eta}g_{n}^{(-)}h_{n}^{(+)}\Big)
\times {\rm id}.
\eea
The expressions of the  $L$-operators (\ref{L-operator}) and (\ref{L-2}) allows us to derive  their quasi-periodicities
\bea
&&L_{n}(u+i\pi)=-\s^z\,L_{n}(u)\,\s^z, \label{L-quasi-periodic}
\eea
\bea
&&\hat{L}_{n}(u+i\pi)=-\s^z\,\hat{L}_{n}(u)\,\s^z. \label{L2-quasi-periodic}
\eea
The quasi-periodicity of K-matrices (\ref{K-periodicity}) enables us to obtain the associated periodicity property of the transfer matrix $t(u)$
\bea
t(u+i\pi)=t(u).\label{transfer-periodic}
\eea
The above relation implies that the transfer matrix $t(u)$ can be expressed in terms of $e^{2u}$
as a Laurent polynomial of the form
\bea
t(u)=e^{2(N+2)u}t_{N+2}+e^{2(N+1)u}t_{N+1}+\cdots+e^{-2(N+2)u}t_{-(N+2)},\label{Expansion-1}
\eea where $\{t_n|n=N+2,N+1,\cdots,-(N+2)\}$ form the $2N+5$ conserved charges. In particular, $t_{N+2}$ and $t_{-(N+2)}$ are
given by
\bea
&&t_{N+2}=-\frac{1}{4}\,e^{(N+2)\eta}\lt\{e^{\theta_+-\theta_-}\,F^{(+)}F^{(-)}+e^{-\theta_++\theta_-}\,D^{(+)}D^{(-)}\rt\},\no\\
&&t_{-(N+2)}=-\frac{1}{4}\,e^{-(N+2)\eta}\lt\{e^{\theta_+-\theta_-}\,F^{(+)}F^{(-)}+e^{-\theta_++\theta_-}\,D^{(+)}D^{(-)}\rt\},\no
\eea where the  constants $D^{(\pm)}$ and $F^{(\pm)}$ are determined by (\ref{Constant-1}).

Following the method in \cite{Beh96,Yan08-e} and using the crossing relation of the
L-matrix (\ref{crosing-unitarity}) and the explicit expressions of
the K-matrices (\ref{K_-(u)}) and (\ref{K +(u)}), we verify that
the corresponding transfer matrix $t(u)$ satisfies the following
crossing relation
\bea
t(-u-\eta)&=&t(u).\label{crosing-opertaor}
\eea


We can define the average value ${\cal{O}}(u)$ of the matrix elements of the monodromy matrices $T(u)$ and $\hat{T}(u)$ (or the
$L$-operators $L_n(u)$ and the $\hat{L}$-operators $\hat{L}_n(u)$) by using the averaging procedure \cite{Tar92}:
\bea
 {\cal{O}}(u)=\prod_{m=1}^p\,O(u-m\eta),\no
\eea where the operator $O(u)$ can be either $\{{\bf A}(u),\,{\bf
B}(u),\,{\bf C}(u),\,{\bf D}(u),\,{\bf \hat{A}}(u),\,{\bf \hat{B}}(u),\,{\bf \hat{ C}}(u),\,{\bf \hat{D}}(u)\}$ or  $\{A_n(u), B_n(u), C_n(u), D_n(u), \hat{A}_n(u), \hat{B}_n(u), \hat{C}_n(u), \hat{D}_n(u)$ $| n=1, \cdots, N\}$. It was shown in Ref. \cite{Tar92} that
\bea
&&{\cal{T}}(u)=\left(\begin{array}{cc}
                           {\bf {\cal{A}}}(u) & {\bf {\cal{B}}}(u) \\[6pt]
                           {\bf {\cal{C}}}(u) & {\bf {\cal{D}}}(u)
                         \end{array}
\right)={\cal{L}}_N(u)\,{\cal{L}}_{N-1}(u)\,\cdots {\cal{L}}_1(u),\label{Average-1}
\eea
\bea
&&{{\hat{\cal{T}}}}(u)=\left(\begin{array}{cc}
                           {\bf {\hat{\cal{A}}}}(u) & {\bf {\hat{\cal{B}}}}(u) \\[6pt]
                           {\bf {\hat{\cal{C}}}}(u) & {\bf {\hat{\cal{D}}}}(u)
                         \end{array}
\right)={\hat{\cal{L}}}_1(u)\,{\hat{\cal{L}}}_{2}(u)\,\cdots {\hat{\cal{L}}}_N(u),\label{Average-2}
\eea
and the average values of each $L$-operator and $\hat{L}$-operator are given by
\bea
 {\cal{L}}_n(u)&=&\left(\begin{array}{cc}
                           {{\cal{A}}}_n(u) & { {\cal{B}}}_n(u) \\[6pt]
                           {{\cal{C}}}_n(u) & {{\cal{D}}}_n(u)
                         \end{array}
\right)\no\\[6pt]
&=&
\left(
\begin{array}{cc}
  e^{pu}\{d^{(+)}_n\}^p+e^{-pu}\{d^{(-)}_n\}^p & \{g^{(+)}_n\}^p+\{g^{(-)}_n\}^p \\[6pt]
  \{h^{(+)}_n\}^p+\{h^{(-)}_n\}^p & e^{pu}\{f^{(+)}_n\}^p+e^{-pu}\{f^{(-)}_n\}^p
\end{array}\right), \label{Average-3}
\eea
\bea
 {\hat{\cal{L}}}_n(u)&=&\left(\begin{array}{cc}
                           {{\hat{\cal{A}}}}_n(u) & {{\hat{\cal{B}}}}_n(u) \\[6pt]
                           {{\hat{\cal{C}}}}_n(u) & {{\hat{\cal{D}}}}_n(u)
                         \end{array}
\right)\no\\[6pt]
&=&
\left(
\begin{array}{cc}
  e^{pu}\{f^{(-)}_n\}^p+e^{-pu}\{f^{(+)}_n\}^p & -\{g^{(+)}_n\}^p-\{g^{(-)}_n\}^p \\[6pt]
  -\{h^{(+)}_n\}^p-\{h^{(-)}_n\}^p & e^{pu}\{d^{(-)}_n\}^p+e^{-pu}\{d^{(+)}_n\}^p
\end{array}\right), \label{Average-4}
\eea
with $n=1, \cdots, N$.
Note that the average values of
the matrix elements are Laurent polynomials of $e^{pu}$, which
implies \bea
&&{\cal{T}}(u+\eta)={\cal{T}}(u),\quad {\cal{L}}_n(u+\eta)={\cal{L}}_n(u),\quad n=1,\cdots,N, \label{P-periodic}\\
&&{\hat{\cal{T}}}(u+\eta)={\hat{\cal{T}}}(u),\quad {\hat{\cal{L}}}_n(u+\eta)={\hat{\cal{L}}}_n(u),\quad n=1,\cdots,N, \label{P-periodic1}\\
&&\lim_{u\rightarrow\pm \infty}{\bf {\cal{A}}}(u)=e^{\pm pNu}\lt\{D^{(\pm )}\rt\}^p,\label{A-asymp}\\
&&\lim_{u\rightarrow\pm \infty}{\bf {\hat{\cal{A}}}}(u)=e^{\pm pNu}\lt\{F^{(\mp )}\rt\}^p,\label{A-asymp2}\\
&&\lim_{u\rightarrow\pm \infty}{\bf {\cal{D}}}(u)=e^{\pm pNu}\lt\{F^{(\pm )}\rt\}^p,\label{D-asymp}\\
&&\lim_{u\rightarrow\pm \infty}{\bf {\hat{\cal{D}}}}(u)=e^{\pm pNu}\lt\{D^{(\mp )}\rt\}^p.\label{D-asymp2}
\eea

\subsection{Fusion hierarchy}
The main tool adopted in this paper  to solve the open $\tau_2$-model is the so-called fusion technique, by which high-dimensional representations can be obtained from the low-dimensional ones. The fusion technique was first developed in Refs. \cite{Kulish1981, Kulish1982, Kirillov1982} for $R-$matrices, and then generalised for $K-$matrices in Refs. \cite{Mezincescu1990, Mezincescu1992, Zhou1996, Inami1996}. In recent years, this technique has been extensively used in solving lots of integrable models \cite{Cao2014, Cao2015}.  Following the procedure in Ref. \cite{Kulish1982}, we introduce the projectors
\bea
P^{(+)}_{1\cdots m}=\frac{1}{m!}\sum_{\sigma\in S_m}\mathcal{P}_{\sigma},
\eea
where $S_m$ is the permutation group of  $m$ indices, and $\mathcal{P}_\sigma$ is the permutation operator in the tensor space $\otimes_{k=1}^{m}\mathcal{C}^2$. For instance,
\bea
P_{12}^{(+)}&=&\frac{1}{2}(1+\mathcal{P}_{12}),\no\\[6pt]
P_{123}^{(+)}&=&\frac{1}{6}(1+\mathcal{P}_{23}\mathcal{P}_{12}+\mathcal{P}_{12}\mathcal{P}_{23}+\mathcal{P}_{12}+\mathcal{P}_{23}+\mathcal{P}_{13}).\no
\eea
The fused spin-$j$ $K^-$-matrix is given by \cite{Mezincescu1992, Zhou1996}
\bea
K^{-(j)}_{\{a\}}(u)&=&P^{(+)}_{a_1,\cdots, a_{2j}}\prod^{2j}_{k=1}\Big\{\Big[\prod_{l=1}^{k-1}R_{a_l,a_k}(2u+(k+l-2j-1)\eta)\Big]\no\\[6pt]
&\times& K_{a_k}^{-(\frac{1}{2})}(u+(k-j-\frac{1}{2})\eta)\Big\}P^{(+)}_{a_1,\cdots,a_2j},\label{fusedk}
\eea
where $\{a\}\equiv\{a_1,\cdots, a_{2j}\}$ and $K^{-(\frac{1}{2})}(u)=K^-(u)$.
The fused spin-$j$ $K^+$-matrix is given by duality
\bea
K^{+(j)}_{\{a\}}(u)=\frac{1}{f^{(j)}(u)}K^{-(j)}_{\{a\}}(-u-\eta)\Big{|}_{(\alpha_-, \beta_-, \theta_-)\rightarrow(-\alpha_+, -\beta_+, \theta_+)},\label{K^+(u)}\label{fused-K2}
\eea
where the normalization factor $f^{(j)}(u)$ is
\bea
f^{(j)}(u)&=&\prod_{l=1}^{2j-1}\prod_{k=1}^{l}[-\rho(2u+(l+k+1-2j)\eta)],\no
\eea
with
\bea
\rho(u)&=&\sinh(u-\eta)\sinh(u+\eta).
\eea
The fused (boundary) matrices satisfy the generalized (boundary) Yang-Baxter equations \cite{Mezincescu1992, Zhou1996}.

We introduce further the fused spin-$j$ monodromy matrices $T^{(j)}_{\{a\}}(u)$ and $\hat{T}^{(j)}_{\{a\}}(u)$  in terms of the fundamental monodromy matrices  $T^{(\frac{1}{2})}(u)=T(u)$  and $\hat{T}^{(\frac{1}{2})}(u)=\hat{T}(u)$ as follows:
\bea
T^{(j)}_{\{a\}}(u)&=&P^{(+)}_{1,\cdots,2j}T_{1}^{(\frac{1}{2})}(u-(j-\frac{1}{2})\eta)T_{2}^{(\frac{1}{2})}(u-(j-\frac{1}{2})\eta+\eta)\no\\[6pt]
&\times&\cdots T_{2j}^{(\frac{1}{2})}(u+(j-\frac{1}{2})\eta)P^{(+)}_{1,\cdots,2j},\label{T1-fused-function}
\eea
\bea
\hat{T}^{(j)}_{\{a\}}(u)&=&P^{(+)}_{1,\cdots,2j}\hat{T}_{1}^{(\frac{1}{2})}(u-(j-\frac{1}{2})\eta)\hat{T}_{2}^{(\frac{1}{2})}(u-(j-\frac{1}{2})\eta+\eta)\no\\[6pt]
&\times&\cdots \hat{T}_{2j}^{(\frac{1}{2})}(u+(j-\frac{1}{2})\eta)P^{(+)}_{1,\cdots,2j}.\label{T2-fused-function}
\eea
The fused transfer matrices $t^{(j)}(u)$ which correspond to a spin-$j$ auxiliary space can be constructed by the fused monodromy matrices and $K$-matrices as
\bea
t^{(j)}(u)=tr_{\{a\}}\Big\{K^{+(j)}_{\{a\}}(u)T_{\{a\}}^{(j)}(u)K^{-(j)}_{\{a\}}(u)\hat{T}_{\{a\}}^{(j)}(u)\Big\}\label{tj-function}.
\eea

The double-row transfer matrix $t(u)$ given by (\ref{transfer}) corresponds to the fundamental case $j=\frac{1}{2}$; that is $t^{(\frac{1}{2})}(u)=t(u)$. Also, the fused transfer matrices constitute commutative families
\bea
[t^{(j)}(u),\,t^{(j')}(v)]=0,\quad j,j'\in \frac
12, 1, \frac{3}{2}, \cdots\label{Commut}.
\eea
These transfer matrices also satisfy the so-called fusion hierarchy \cite{Mezincescu1990, Mezincescu1992, Zhou1996, Fra07}
\bea
 t^{(\frac{1}{2})}(u)\, t^{(j-\frac{1}{2})}(u- j\eta)&=&
t^{(j)}(u-(j-\frac{1}{2})\eta) + \delta(u)\,
t^{(j-1)}(u-(j+\frac{1}{2})\eta),\no\\[6pt]
j &=&\frac 12, 1, \frac{3}{2}, \cdots, \label{Hier-1} \eea
with the conventions $t^{(-\frac{1}{2})}(u)=0$ and
$t^{(0)}={\rm id}$.
The coefficient $\delta(u)$, the so-called quantum determinant \cite{Kulish1981, Ize81, PP1982, AG1982}, is given by
\bea
\delta(u)=-\nu(u){\mathrm{Det}}_{q}\{K^+(u)\}~{\mathrm{Det}}_{q}\{T(u)\}~{\mathrm{Det}}_{q}\{K^-(u)\}~{\mathrm{ Det}}_{q}\{\hat{T}(u)\},\label{det}
\eea where
\bea
\nu(u)=\frac{1}{\sinh(2u+\eta)\sinh(2u-\eta)},
\eea
\bea
\mathrm{Det}_q\{T(u)\}&=&tr_{12}\{P^{-}_{12}T_1(u)T_2(u+\eta)\}=\prod_{n=1}^N\,
{\mathrm{Det}}_{q}\{L_n(u)\},\label{t-function}\\[6pt]
{\mathrm{Det}}_{q}\{L_n(u)\}&=&e^{2u-\eta}d_{n}^{(+)}f_{n}^{(+)}+e^{-2u+\eta}d_{n}^{(-)}f_{n}^{(-)}
-e^{\eta}g_{n}^{(+)}h_{n}^{(-)}-e^{-\eta}g_{n}^{(-)}h_{n}^{(+)},\label{det-t}
\eea
\bea
{\rm Det}_q\{\hat{T}(u)\}&=&tr\{P^{-}_{12}\hat{T}_1(u)\hat{T}_2(u+\eta)\}=\prod_{n=1}^N\,
{\rm Det}_q\{\hat{L}_n(u)\},\label{t1-function}\\[6pt]
{\mathrm{Det}}_{q}\{\hat{L}_n(u)\}&=&e^{-2u-\eta}d_{n}^{(+)}f_{n}^{(+)}+e^{2u+\eta}d_{n}^{(-)}f_{n}^{(-)}
-e^{\eta}g_{n}^{(+)}h_{n}^{(-)}-e^{-\eta}g_{n}^{(-)}h_{n}^{(+)},\label{det-t-1}\\[6pt]
\mathrm{Det}_{q}\{K^-(u)\}&=&tr_{12}\{P_{12}K^{-}_{1}(u)R_{12}(2u+\eta)K^{-}_2(u+\eta)\}\no\\
&=&-2^2\sinh(2u-2\eta)\sinh(u+\alpha_-)\sinh(u-\alpha_-)\no\\
&&\quad\quad\times\cosh(u+\beta_-)\cosh(u-\beta_-),\\
{\mathrm{Det}}_{q}\{K^+(u)\}&=&tr_{12}\{P_{12}K^{+}_{2}(u+\eta)R_{12}(-2u-3\eta)K^{+}_2(u)\}\no\\
&=&\mathrm{Det}_q\{K^-(-u-2\eta)\}|_{(\alpha_{-}, \beta_-, \theta_-)\rightarrow(-\alpha_+, -\beta_+, \theta_+)}\no\\
&=&2^2\sinh(2u+2\eta)\sinh(u+\alpha_+)\sinh(u-\alpha_+)\no\\
&&\quad \quad\times\cosh(u+\beta_+)\cosh(u-\beta_+).\label{det2}
\eea

Let us introduce the functions $a(u)$ and $d(u)$ as follows:
\bea
a(u)&=&-2^2\,\frac{\sinh(2u+2\eta)}{\sinh(2u+\eta)}\,\sinh(u-\alpha_-)\,\sinh(u-\alpha_+)\no\\
&&\quad\quad\times\cosh(u-\beta_-)\,\cosh(u-\beta_+)\,\bar{A}(u),\label{a2-function}\\
d(u)&=&a(-u-\eta),\label{d2-function}
\eea
where
\bea
\bar{A}(u)=e^{-N\eta}G^{(-)}H^{(+)}\prod_{n=1}^N\lt(e^u-e^{-u
   +2\eta}\,\frac{d^{(-)}_nf^{(-)}_n}{g^{(-)}_nh^{(+)}_n}\rt)\,
   \lt(e^{-u}\,\frac{d^{(+)}_nf^{(+)}_n}{g^{(-)}_nh^{(+)}_n}-e^u\rt).\label{A-function}
\eea
Similar as the definitions (\ref{Constant-1}), the  constants  $G^{(\pm)}$ and
$H^{(\pm)}$ are
\bea G^{(\pm)}=\prod_{n=1}^Ng_n^{(\pm)},\quad
H^{(\pm)}=\prod_{n=1}^Nh_n^{(\pm)}. \label{Constant-2}
\eea
Then it is easy to check that the quantum determinant (\ref{det}) can be expressed in terms of the above functions as
\bea
\delta(u)&=&a(u)d(u-\eta).\label{delta-function}\eea

Similar to the $\tau_2$-model with the periodic condition \cite{Xu15}, we can also use the recursive relation (\ref{Hier-1}) and the coefficient function (\ref{delta-function}) to express the fused transfer matrix $t^{(j)}(u)$ in terms of the fundamental one $t^{(\frac{1}{2})}(u)$ with a $2j$-order functional relation which can be expressed as the
determinant of some $2j \times 2j$ matrix \cite{Bzahanov1989}, namely,
{\small \bea
&&\hspace{-1.4truecm}t^{(j)}(u)\hspace{-0.092truecm}=\hspace{-0.092truecm}\lt|\hspace{-0.12truecm}
\begin{array}{ccccc}
                           t(u\hspace{-0.08truecm}+\hspace{-0.08truecm}(j\hspace{-0.08truecm}-\hspace{-0.08truecm}\frac{1}{2})\eta)
                           &\hspace{-0.08truecm}-\hspace{-0.08truecm}a(u\hspace{-0.08truecm}+\hspace{-0.08truecm}(j\hspace{-0.08truecm}-\hspace{-0.08truecm}\frac{1}{2})\eta)
                           &&&\\[6pt]
                           \hspace{-0.08truecm}-\hspace{-0.08truecm}d(u\hspace{-0.08truecm}+\hspace{-0.08truecm}(j\hspace{-0.08truecm}-\hspace{-0.08truecm}\frac{3}{2})\eta) &t(u\hspace{-0.08truecm}+\hspace{-0.08truecm}(j\hspace{-0.08truecm}-\hspace{-0.08truecm}\frac{3}{2})\eta)
                           &\hspace{-0.08truecm}-\hspace{-0.08truecm}a(u\hspace{-0.08truecm}+\hspace{-0.08truecm}(j\hspace{-0.08truecm}-\hspace{-0.08truecm}\frac{3}{2})\eta)
                           &&\\[6pt]
                           &\ddots&&&\\[6pt]
                           &&\cdots&&\\[6pt]
                           &&&\ddots&\\[6pt]
                           &&\hspace{-0.08truecm}-\hspace{-0.08truecm}d(u\hspace{-0.08truecm}-\hspace{-0.08truecm}(j\hspace{-0.08truecm}+\hspace{-0.08truecm}\frac{1}{2})\eta)
                           &t(u\hspace{-0.08truecm}-\hspace{-0.08truecm}(j\hspace{-0.08truecm}+\hspace{-0.08truecm}\frac{1}{2})\eta)
                           &\hspace{-0.08truecm}-\hspace{-0.08truecm}a(u\hspace{-0.08truecm}-\hspace{-0.08truecm}(j\hspace{-0.08truecm}+\hspace{-0.08truecm}\frac{1}{2})\eta)\\[4pt]
                           &&&\hspace{-0.08truecm}-\hspace{-0.08truecm}d(u\hspace{-0.08truecm}-\hspace{-0.08truecm}(j\hspace{-0.08truecm}-\hspace{-0.08truecm}\frac{1}{2})\eta)
                           &t(u\hspace{-0.08truecm}-\hspace{-0.08truecm}(j\hspace{-0.08truecm}-\hspace{-0.08truecm}\frac{1}{2})\eta)
                         \end{array}
\hspace{-0.12truecm}\rt|,\no\\[12pt]
&&\quad\quad j=\frac{1}{2},1,\frac{3}{2},\cdots.\label{Dert-rep}
\eea}

\section{Truncation identity}
\setcounter{equation}{0}

We now proceed to formulate the desired operator identities to determine the spectrum of the transfer matrix $t(u)$ given by (\ref{transfer}). For this purpose, we first derive separate truncation identities for the monodromy matrices and $K$-matrices.
We recall that the fusion approach described in the previous section. When the crossing parameters $\eta$ takes the special values $\eta=\frac{2i\pi}{p}$, one can find that the spin-$\frac{p}{2}$ fused monodromy matrices  mentioned in (\ref{T1-fused-function}), (\ref{T2-fused-function}), all take the block-lower triangular forms \cite{Tar92}
\bea
T^{(\frac{p}{2})}(u)&=&
\left(
  \begin{array}{ccc}
   {\bf {\cal{A}}}(u) & {\bf {\cal{B}}}(u) & 0 \\[6pt]
    {\bf {\cal{C}}}(u) & {\bf {\cal{D}}}(u) & 0 \\[6pt]
   g(u) & h(u) & {\mathrm{Det}_{q}\{T(u-(\frac{p-1}{2})\eta)}\}F T^{(\frac{p}{2}-1)}(u) F^{-1} \label{t-fused}\\
  \end{array}
\right),\\[8pt]
\hat{T}^{(\frac{p}{2})}(u)&=&
\left(
  \begin{array}{ccc}
   {\bf {\hat{\cal{A}}}}(u) & {\bf {\hat{\cal{B}}}}(u) & 0 \\[6pt]
    {\bf {\hat{\cal{C}}}}(u) & {\bf {\hat{\cal{D}}}}(u) & 0 \\[6pt]
   \hat{g}(u) & \hat{h}(u) & {\mathrm{Det}_{q}\{\hat{T}(u-(\frac{p-1}{2})\eta)}\}F \hat{T}^{(\frac{p}{2}-1)}(u) F^{-1} \label{tt-fused}\\
  \end{array}
\right),
\eea
where
\bea
F&=&M\otimes\sigma_z,\label{f-matrix}
\eea
with
\bea
M&=&\left(
    \begin{array}{ccccc}
      1 & 0 & 0 & 0 &0  \\[5pt]
      0 & [2]_q & 0 & 0 & 0 \\[5pt]
      0 & 0 & [3]_q & 0 & 0 \\[5pt]
      0 & 0 & 0 &\ddots & 0 \\[5pt]
      0 & 0 & 0 & 0 & [\frac{p-1}{2}]_q \\
    \end{array}
  \right),
\eea
and \bea
[x]=\frac{q^x-q^{-x}}{q-q^{-1}},~~~~~~q=e^{-\eta}.
\eea
Here $\{g(u),h(u),\hat{g}(u),\hat{h}(u)\}$ are some operators, which are irrelevant to the transfer matrices.
Moreover, through tedious calculations, it is found  that the general non-diagonal boundary fused
$K^-$-matrices (\ref{fusedk}) for $\eta=\frac{2i\pi}{p}$ and $j=\frac{p}{2}$ take the following form like (\ref{t-fused})
\bea
K^{-(\frac{p}{2})}(u)= \mu^{(\frac{p}{2})}(u)
\left(
  \begin{array}{ccc}
  \mathcal{K}_{11}^{-(\frac{p}{2})}(u) & \mathcal{K}_{12}^{-(\frac{p}{2})}(u)  & 0 \\[6pt]
  \mathcal{K}_{21}^{-(\frac{p}{2})}(u) &\mathcal{ K}_{22}^{-(\frac{p}{2})}(u)  & 0 \\[6pt]
   k_3(u) & k_4(u) & K_{33}^{-(\frac{p}{2})}(u)\label{tk-fused}\\
  \end{array}
\right),
\eea
where the functions
\bea
\mu^{(\frac{p}{2})}(u)&=&\prod_{l=1}^{p-1}\prod_{k=1}^{l}\sinh(2u+(l+k-p+1)\eta),\\\label{m-function}
\mathcal{K}_{11}^{-(\frac{p}{2})}(u)&=&\sum_{l=0}^{[\frac{p}{2}]}c_{p}^{2l}(\sinh\alpha_-\cosh\beta_-)^{p-2l}(\cosh\alpha_-\sinh\beta_-)^{2l}\cosh(pu)\no\\[6pt]
&+&\sum_{l=0}^{[\frac{p}{2}]}c_{p}^{2l+1}(\cosh\alpha_-\sinh\beta_-)^{p-2l-1}(\sinh\alpha_-\cosh\beta_-)^{2l+1}\sinh(pu)\no\\[6pt]
&+& p(\sinh\alpha_-\cosh\beta_-\cosh(pu)+\cosh\alpha_-\sinh\beta_-\sinh(pu)),\label{k11}\\[6pt]
\mathcal{K}_{22}^{-(\frac{p}{2})}(u)&=&\sum_{l=0}^{[\frac{p}{2}]}c_{p}^{2l}(\sinh\alpha_-\cosh\beta_-)^{p-2l}(\cosh\alpha_-\sinh\beta_-)^{2l}\cosh(pu)\no\\[6pt]
&-&\sum_{l=0}^{[\frac{p}{2}]}c_{p}^{2l+1}(\cosh\alpha_-\sinh\beta_-)^{p-2l-1}(\sinh\alpha_-\cosh\beta_-)^{2l+1}\sinh(pu)\no\\[6pt]
&+& p(\sinh\alpha_-\cosh\beta_-\cosh(pu)-\cosh\alpha_-\sinh\beta_-\sinh(pu)),
\eea
and the $(p-1)\times(p-1)$ matrix $K_{33}^{-(\frac{p}{2})}(u)$ is related to the spin-$\frac{p-2}{2}$ $K^-$-matrix by
\bea
K_{33}^{-(\frac{p}{2})}(u)&=&\frac{{\mathrm{Det}}_{q}\{K^{-}(u-(\frac{p-1}{2})\eta)\}}{\sinh[2(u-(\frac{p-1}{2})\eta)+\eta]}F K^{-(\frac{p-2}{2})}(u) F^{-1}.
\eea
The functions $\mathcal{K}_{12}^{-(\frac{p}{2})}(u)$ and $\mathcal{K}_{21}^{-(\frac{p}{2})}(u)$ give contributions to calculating the asymptotic behavior of the eigenvalue of $t(u)$ (\ref{transfer}) and possess the following forms respectively
\bea
\mathcal{K}_{12}^{-(\frac{p}{2})}(u)&=&(\frac{1}{2})^{p-1}e^{p\theta_-}\sinh(2pu),\label{k1-element}\\
\mathcal{K}_{21}^{-(\frac{p}{2})}(u)&=&(\frac{1}{2})^{p-1}e^{-p\theta_-}\sinh(2pu),
\eea
with the matrix $F$ being given by (\ref{f-matrix}). Moreover, $k_3(u)$ and $k_4(u)$ are two $p\times 1$ matrices which
are irrelevant to the associated transfer matrix.

At the same time, the fused $K^+$-matrices are given, in view of Eq. (\ref{fused-K2}) with $\eta=\frac{2i\pi}{p}$ and $j=\frac{p}{2}$, by
\bea
K^{+(\frac{p}{2})}_{\{a\}}(u)=\frac{1}{f^{(\frac{p}{2})}(u)}K^{-(\frac{p}{2})}_{\{a\}}(-u-\eta)\Big{|}_{(\alpha_-, \beta_-, \theta_-)\rightarrow(-\alpha_+, -\beta_+, \theta_+)}.\label{kp-function}
\eea
The explicit expressions of the elements of the fused monodromy matrices and the $K$-matrices for the cases $p=3$ are given in Appendix A.

Hence, we are finally in position to formulate the truncation identity for the fused transfer matrices $t^{(j)}(u)$ defined in (\ref{tj-function}).
Based on the results of (\ref{t-fused}) and (\ref{tt-fused}) for the fused monodromy matrices and those of (\ref{tk-fused}) and (\ref{kp-function}) for the fused $K$-matrices, we obtain
\bea
t^{(\frac{p}{2})}(u)=(\tilde{\mathcal{A}}(u)+\tilde{\mathcal{D}}(u))\times {\rm id}+\delta\Big(u-(\frac{p-1}{2})\eta\Big)t^{(\frac{p-2}{2})}(u),\label{transfer-fused}
\eea
where the coefficients $\tilde{\mathcal{A}}(u)$ and $\tilde{\mathcal{D}}(u)$ are given by\footnote{The average values $\{\mathcal{A}(u),\,\mathcal{B}(u),\,\mathcal{C}(u),\,\mathcal{D}(u)\}$ and $\{\mathcal{\hat{A}}(u),\,\mathcal{\hat{B}}(u),\,
\mathcal{\hat{C}}(u),\,\mathcal{\hat{D}}(u)\}$ of the matrix elements of the monodromy matrices $T(u)$ and $\hat{T}(u)$  in (\ref{a-average}) and (\ref{d-average}) can be obtained from the relations in (\ref{Average-1}) and (\ref{Average-2}) which are based on the average values of each $L$-operator and $\hat{L}$-operator given in (\ref{Average-3}) and (\ref{Average-4}).  For some examples of the small sites such as $N=1,2$, we give the explicit expressions of the average value functions and also  discuss some special constraints that allow one to  calculate these average value functions for an arbitrary number of the sites in Appendix B.}.
\bea
\tilde{\mathcal{A}}(u)=[\mathcal{K}_{11}^{+(\frac{p}{2})}(u)\mathcal{A}(u)+\mathcal{K}_{12}^{+(\frac{p}{2})}(u)\mathcal{C}(u)]
[\mathcal{K}_{11}^{-(\frac{p}{2})}(u)\mathcal{\hat{A}}(u)
+\mathcal{K}_{12}^{-(\frac{p}{2})}(u)\mathcal{\hat{C}}(u)]\no\\[6pt]+[\mathcal{K}_{11}^{+(\frac{p}{2})}(u)
\mathcal{B}(u)+\mathcal{K}_{12}^{+(\frac{p}{2})}(u)\mathcal{D}(u)][\mathcal{K}_{21}^{-(\frac{p}{2})}(u)
\mathcal{\hat{A}}(u)+\mathcal{K}_{22}^{-(\frac{p}{2})}(u)\mathcal{\hat{C}}(u)],\label{a-average}\\[6pt]
\tilde{\mathcal{D}}(u)=[\mathcal{K}_{21}^{+(\frac{p}{2})}(u)\mathcal{A}(u)+\mathcal{K}_{22}^{+(\frac{p}{2})}(u)\mathcal{C}(u)]
[\mathcal{K}_{11}^{-(\frac{p}{2})}(u)\mathcal{\hat{B}}(u)
+\mathcal{K}_{12}^{-(\frac{p}{2})}(u)\mathcal{\hat{D}}(u)]\no\\[6pt]
+[\mathcal{K}_{21}^{+(\frac{p}{2})}(u)\mathcal{B}(u)+\mathcal{K}_{22}^{+(\frac{p}{2})}(u)\mathcal{D}(u)][\mathcal{K}_{21}^{-(\frac{p}{2})}(u)\mathcal{\hat{B}}(u)
+\mathcal{K}_{22}^{-(\frac{p}{2})}(u)\mathcal{\hat{D}}(u)].\label{d-average}
\eea
It is remarked that the functions $\mathcal{K}_{11}^{\pm(\frac{p}{2})}(u), \mathcal{K}_{12}^{\pm(\frac{p}{2})}(u), \mathcal{K}_{21}^{\pm(\frac{p}{2})}(u), \mathcal{K}_{22}^{\pm(\frac{p}{2})}(u)$ and the average values of each monodromy matrices are invariant under shifting with $\eta$.

Combining the fusion hierarchy (\ref{Hier-1}) and the closing relation (\ref{transfer-fused}) for $\eta=\frac{2i\pi}{p}$, we arrive at the functional relation for the fundamental transfer matrix straightforward. Here we give an example of the functional relations for $p=3$:
\bea
t(u+\eta)\,t(u)\,t(u-\eta)-\d(u+\eta)\,t(u-\eta)-\d(u)\,t(u+\eta)=\tilde{\mathcal{A}}(u)+\tilde{\mathcal{D}}(u)+\delta(u-\eta)t(u).\no
\eea

\section{Eigenvalues of the fundamental transfer matrix}
\setcounter{equation}{0}

\subsection{Functional relations of eigenvalues}

The commutativity (\ref{Commut}) of the fused transfer matrices $\{t^{(j)}(u)\}$ with different spectral
parameters implies that they have common eigenstates. One can set $|\Psi\rangle$ to be a common eigenstate of these
fused transfer matrices with eigenvalues $\Lambda^{(j)}(u)$, i.e.,
\bea
t^{(j)}(u)|\Psi\rangle =\Lambda^{(j)}(u)|\Psi\rangle.\no
\eea
The quasi-periodicity (\ref{transfer-periodic}) and the cross relation (\ref{crosing-opertaor}) of the transfer
matrix $t(u)$ implies that the corresponding eigenvalue $\L(u)$
satisfies the properties \bea
\L(u+i\pi)=\L(u),\label{Eigen-periodic}\\[6pt]
\Lambda(-u-\eta)=\Lambda(u).\label{cross-relation}
\eea
The asymptotic behavior (\ref{Asymp}) and the special values at $u=0,\frac{i\pi}{2}$ of the transfer matrix $t(u)$ enables us to derive that the corresponding eigenvalue $\L(u)$ have the following functional relations \bea \lim_{u\rightarrow \pm\infty}\L(u)&=&-\frac{1}{4}e^{\pm \{(2N+4)u+(N+2)\eta\}}\lt\{e^{\theta_+-\theta_-}\,F^{(+)}F^{(-)}+e^{-\theta_++\theta_-}\,D^{(+)}D^{(-)}\rt\},\label{Eigen-Asymp}\\[6pt]
\Lambda(0)&=&-2^3\sinh\a_-\cosh\b_-\sinh\a_+\cosh\b_+\cosh\eta\,\no\\[6pt]
&&\times
\prod_{n=1}^N\Big(e^{-\eta}d_{n}^{(+)}f_{n}^{(+)}+e^{\eta}d_{n}^{(-)}f_{n}^{(-)}-e^{\eta}g_{n}^{(+)}h_{n}^{(-)}-e^{-\eta}g_{n}^{(-)}h_{n}^{(+)}\Big),\\[6pt]
\Lambda(\frac{i\pi}{2})&=&-2^3\cosh\a_-\sinh\b_-\cosh\a_+\sinh\b_+\cosh\eta\,\no\\[6pt]
&&\times
\prod_{n=1}^N\Big(e^{-\eta}d_{n}^{(+)}f_{n}^{(+)}+e^{\eta}d_{n}^{(-)}f_{n}^{(-)}+e^{\eta}g_{n}^{(+)}h_{n}^{(-)}+e^{-\eta}g_{n}^{(-)}h_{n}^{(+)}\Big).\label{special-values2}
\eea
The analyticity of the $L$-operator (\ref{L-operator}), the
quasi-periodicity (\ref{Eigen-periodic}) and the asympotic behavior (\ref{Eigen-Asymp}) of the eigenvalue
give rise to that $\Lambda(u)$ possesses the following
analytical property \bea \L(u) \mbox{, as a function of $e^u$, is
a Laurent polynomial of degree $2N+4$ like}~
(\ref{Expansion-1}).\label{Eigenvalue-Anal} \eea
According to the fusion hierarchy relation (\ref{Hier-1}) and the determinant
representation (\ref{Dert-rep}) of the fused transfer matrices, the eigenvalues $\Lambda^{(j)}(u)$ give some similar representation in
terms of the fundamental one $\L(u)=\Lambda^{(\frac{1}{2})}(u)$
{\small \bea
&&\hspace{-1.4truecm}\L^{(j)}(u)\hspace{-0.092truecm}=\hspace{-0.092truecm}\lt|\hspace{-0.12truecm}
\begin{array}{ccccc}
                           \L(u\hspace{-0.08truecm}+\hspace{-0.08truecm}(j\hspace{-0.08truecm}-\hspace{-0.08truecm}\frac{1}{2})\eta)
                           &\hspace{-0.08truecm}-\hspace{-0.08truecm}a(u\hspace{-0.08truecm}+\hspace{-0.08truecm}(j\hspace{-0.08truecm}-\hspace{-0.08truecm}\frac{1}{2})\eta)
                           &&&\\[6pt]
                           \hspace{-0.08truecm}-\hspace{-0.08truecm}d(u\hspace{-0.08truecm}+\hspace{-0.08truecm}(j\hspace{-0.08truecm}-\hspace{-0.08truecm}\frac{3}{2})\eta) &\L(u\hspace{-0.08truecm}+\hspace{-0.08truecm}(j\hspace{-0.08truecm}-\hspace{-0.08truecm}\frac{3}{2})\eta)
                           &\hspace{-0.08truecm}-\hspace{-0.08truecm}a(u\hspace{-0.08truecm}+\hspace{-0.08truecm}(j\hspace{-0.08truecm}-\hspace{-0.08truecm}\frac{3}{2})\eta)
                           &&\\[6pt]
                           &\ddots&&&\\[6pt]
                           &&\cdots&&\\[6pt]
                           &&&\ddots&\\[6pt]
                           &&\hspace{-0.08truecm}-\hspace{-0.08truecm}d(u\hspace{-0.08truecm}-\hspace{-0.08truecm}(j\hspace{-0.08truecm}+\hspace{-0.08truecm}\frac{1}{2})\eta)
                           &\L(u\hspace{-0.08truecm}-\hspace{-0.08truecm}(j\hspace{-0.08truecm}+\hspace{-0.08truecm}\frac{1}{2})\eta)
                           &\hspace{-0.08truecm}-\hspace{-0.08truecm}a(u\hspace{-0.08truecm}-\hspace{-0.08truecm}(j\hspace{-0.08truecm}+\hspace{-0.08truecm}\frac{1}{2})\eta)\\[4pt]
                           &&&\hspace{-0.08truecm}-\hspace{-0.08truecm}d(u\hspace{-0.08truecm}-\hspace{-0.08truecm}(j\hspace{-0.08truecm}-\hspace{-0.08truecm}\frac{1}{2})\eta)
                           &\L(u\hspace{-0.08truecm}-\hspace{-0.08truecm}(j\hspace{-0.08truecm}-\hspace{-0.08truecm}\frac{1}{2})\eta)
                         \end{array}
\hspace{-0.12truecm}\rt|,\no\\[8pt]
&&\quad\quad j=\frac{1}{2},1,\frac{3}{2},\cdots,\label{Eigenvlue-2}
\eea} where the functions $a(u)$ and $d(u)$ are given by
(\ref{a2-function}) and (\ref{d2-function}).
The truncation identity (\ref{transfer-fused}) of the spin-$\frac{p}{2}$ transfer matrix leads to the fact that the corresponding
eigenvalue $\L^{(\frac{p}{2})}(u)$ satisfies the relation
\bea
\Lambda^{(\frac{p}{2})}(u)=\tilde{\mathcal{A}}(u)+\tilde{\mathcal{D}}(u)+\delta\Big(u-(\frac{p-1}{2})\eta\Big)\Lambda^{(\frac{p-2}{2})}(u),\label{eigenvalue-fused}
\eea
where the functions $\tilde{\mathcal{A}}(u)$ and $\tilde{\mathcal{D}}(u)$ are given by (\ref{a-average})-(\ref{d-average}). For example, the functional relation of the eigenvalue for $p=3$ is
\bea
\Lambda(u+\eta)\,\Lambda(u)\,\Lambda(u-\eta)-\d(u+\eta)\,\Lambda(u-\eta)-\d(u)\,
\Lambda(u+\eta)=\tilde{\mathcal{A}}(u)+\tilde{\mathcal{D}}(u)+\delta(u-\eta)\Lambda(u).\no
\eea

It is believed \cite{Baxter04, Bazhanov90, Geh06} that the relations (\ref{Eigen-periodic})-(\ref{special-values2}),
the analytic property (\ref{Eigenvalue-Anal}) and
the truncation identity (\ref{eigenvalue-fused}) allow us to completely determine the eigenvalues $\L(u)$
of the fundamental transfer matrix $t(u)$ given by (\ref{transfer}).

\subsection{T-Q relation}
\subsubsection{Generic case}

Following the ODBA method \cite{yu15} and the method developed in \cite{Xu15}, we can express eigenvalues of $t(u)$ in terms of
some inhomogeneous $T-Q$ relation
\bea
\L(u)=a(u)\frac{Q(u-\eta)}{Q(u)}+d(u)\frac{Q(u+\eta)}{Q(u)}+2^{(1-p)2N-4p+2}c\,\frac{\sinh(2u)\sinh(2u+2\eta)F(u)}{Q(u)},\label{T-Q-relation}
\eea where the
functions $a(u)$ and $d(u)$ are given by (\ref{a2-function}) and
(\ref{d2-function}), and the constant $c$ is uniquely determined by the inhomogeneous parameters and boundary parameters,
\bea
&&(\frac{1}{2})^{2p}c\,\Big\{e^{p(\theta_+-\theta_-)}\{F^{(+)}F^{(-)}\}^{p}+e^{-p(\theta_+-\theta_-)}\{D^{(+)}D^{(-)}\}^{p}-(-1)^{N}e^{-p(\alpha_++\beta_++\alpha_-+\beta_-)}
\no\\[6pt]
&&~~~~\times\{G^{(-)}H^{(+)}\}^{p}-(-1)^{N}e^{p(\alpha_++\beta_++\alpha_-+\beta_-)}\{G^{(+)}H^{(-)}\}^{p}\Big\}=\frac{1}{4}\Big\{e^{\theta_+-\theta_-}F^{(+)}F^{(-)}\no\\[6pt]
&&\qquad\qquad+e^{-(\theta_+-\theta_-)}D^{(+)}D^{(-)}-(-1)^{N}e^{-(\alpha_++\beta_++\alpha_-+\beta_-)}\{G^{(-)}H^{(+)}\}e^{-\eta}\no\\[6pt]
&&\qquad\qquad-(-1)^{N}e^{\alpha_++\beta_++\alpha_-+\beta_-}\{G^{(+)}H^{(-)}\}e^{\eta}\Big\}.\label{c-Constant}
\eea
The trigonometric polynomial $Q(u)$ is parameterized by $(p-1)N+2p$ Bethe roots $\{\lambda_j\}$
\bea
Q(u)=\prod_{j=1}^{(p-1)N+2p}\sinh(u-\l_j)\sinh(u+\lambda_j+\eta),\label{Q-function}
\eea
which will be specified by the associated BAEs (\ref{BAE-1}) below.
The function $F(u)$ which is a Laurent polynomial of degree $p(2N+4)$ is given by \bea
&&F(u)={\bf {\cal{\tilde{A}}}}(u)+{\bf {\cal{\tilde{D}}}}(u)-{\bf \bar {\cal{A}}}(u)-{\bf \bar {\cal{D}}}(u), \label{F-fuction}\\
&&{\bf \bar {\cal{A}}}(u)=\prod_{m=1}^p a(u-m\eta),\quad {\bf \bar {\cal{D}}}(u)=\prod_{m=1}^p d(u-m\eta),\label{Bar-a-d-functions}
\eea
where the functions ${\bf {\cal{\tilde{A}}}}(u)$ and ${\bf {\cal{\tilde{D}}}}(u)$  are given in (\ref{a-average}) and (\ref{d-average}) respectively.

According to the relations (\ref{P-periodic})-(\ref{P-periodic1}), the definitions (\ref{Bar-a-d-functions}), (\ref{a-average}) and (\ref{d-average}) and the explicit expression (\ref{k11})-(\ref{kp-function}) of the elements of the $K$-matrices,
the function $F(u)$ can be reduced as a Laurent polynomial of $e^{pu}$ with a degree $2N+4$ (i.e. there are only $2N+5$ non-vanishing coefficients), namely,
\bea
F(u)=\sum_{l=0}^{2N+4}F^{(2p(N+2-l))}(\{d_{n}^{\pm}, f_{n}^{\pm}, g_{n}^{\pm}, h_{n}^{\pm}, d_{n}^{\pm}, \alpha_{\pm}, \beta_{\pm}, \theta_{\pm}\})e^{p(2N+4-2l)u},\label{F-function}
\eea
where the $2N+5$ coefficients $\{F^{(2p(N+2-l))}|l=0,1,\cdots,2N+4\}$ are polynomial of the inhomogeneity parameters $\{d_{n}^{\pm}, f_{n}^{\pm}, g_{n}^{\pm}, h_{n}^{\pm}|n=1,2,\cdots, N\}$ and the boundary parameters $\{\alpha_{\pm}, \beta_{\pm}, \theta_{\pm}\}$,
and that we can easily prove that the function $F(u)$ holds the crossing property
\bea
F(-u-\eta)=F(u).\label{f-function}
\eea
The fact that the constant $c$ satisfies the relation (\ref{c-Constant}) ensures that $\Lambda(u)$ given by (\ref{T-Q-relation})
matches  the asymptotic behavior (\ref{Eigen-Asymp}).  The $(p-1)N+2p$ parameters $\{\l_j|j=1,\cdots,(p-1)N+2p\}$ satisfy the
associated Bethe Ansatz equations (BAEs)
\bea
 && a(\l_j)Q(\l_j - \eta) +d(\l_j)Q(\l_j + \eta) +\, 2^{2(1-p)N-4p+2}c\,\sinh(2\lambda_j) \no \\[6pt]
     && ~~~~~~\qquad\times\sinh(2\lambda_j+2\eta) F(\l_j)=0,\quad j=1,\cdots,(p-1)N+2p,\label{BAE-1}
\eea
which assure that $\Lambda(u)$ given by (\ref{T-Q-relation}) is indeed a trigonometric polynomial of $u$.
It is easy to check that  $\Lambda(u)$ given by the inhomogeneous $T-Q$ relation (\ref{T-Q-relation})
satisfies the the properties (\ref{Eigen-periodic})-(\ref{cross-relation}) and the functional relations (\ref{Eigen-Asymp})-(\ref{special-values2}).
Using the method in the appendix A of \cite{Xu15}, we have checked that the $T-Q$ relation (\ref{T-Q-relation}) also make the functional relation (\ref{eigenvalue-fused}) fulfilled. Therefore, the resulting expression of $\Lambda(u)$ constructed by the inhomogeneous $T-Q$ relation (\ref{T-Q-relation}) is the eigenvalue of the fundamental transfer matrix $t(u)$ of the $\tau_2$-model with generic boundary condition.

\subsubsection{Degenerate case}

The third term of the $T-Q$ relation (\ref{T-Q-relation}) does not  vanish when the generic inhomogeneous parameters
$\{d^{(\pm)}_n,~f^{(\pm)}_n,~g^{(\pm)}_n,~h^{(\pm)}_n
|n=1,\cdots,N\}$ (only obey the constraint (\ref{Constraint-1}) which ensure the integrability of the model)
and the boundary parameters are absence of restriction. Here we consider some special case making the inhomogeneous
term vanishes. When the inhomogeneous parameters
$\{d^{(\pm)}_n,~f^{(\pm)}_n,~g^{(\pm)}_n,~h^{(\pm)}_n
|n=1,\cdots,N\}$ and the boundary parameters $\{\alpha_{\pm},\beta_{\pm}, \theta_{\pm}\}$ obey the following extra constraints besides
(\ref{Constraint-1}):
\bea
&&e^{\theta_+-\theta_-}F^{(+)}F^{(-)}
+e^{-(\theta_+-\theta_-)}D^{(+)}D^{(-)}-(-1)^{N}e^{-(\alpha_++\beta_++\alpha_-+\beta_-)}\{G^{(-)}H^{(+)}\}e^{-\eta}\no\\[6pt]
&&\qquad\qquad -(-1)^{N}e^{\alpha_++\beta_++\alpha_-+\beta_-}\{G^{(+)}H^{(-)}\}e^{\eta}=0,\label{Constraint-2}\\[6pt]
&&F^{(2(N+2-l)p)}(\{d_{n}^{(\pm)}, f_{n}^{(\pm)}, g_{n}^{(\pm)}, h_{n}^{(\pm)}, \alpha_{\pm}, \beta_{\pm}, \theta_{\pm}\})=0,~~~l=1,\cdots,N+2,\label{constraint-9}
\eea
where $D^{(\pm)}$,
$F^{(\pm)}$, $G^{(\pm)}$ and $H^{(\pm)}$ are given by (\ref{Constant-1})
and (\ref{Constant-2}), and each $F^{(2(N+2-l)p)}$ given in
(\ref{F-function}). The corresponding inhomogeneous $T-Q$ relation (\ref{T-Q-relation})
reduces to the conventional one \cite{Bax82}:
\bea
\Lambda(u)=a(u)\frac{\bar{Q}(u-\eta)}{\bar{Q}(u)}
+d(u)\frac{\bar{Q}(u+\eta)}{\bar{Q}(u)},\label{T-Q
relation-2}
\eea
 where the function
$\bar Q(u)$ becomes \cite{Cao1, yu15, Fra07, Nep01, Cao033, Yang06, Cao15} \bea
\bar Q(u)=\prod_{j=1}^{M}\sinh(u-\l_j)\sinh(u+\lambda_j+\eta).\label{Q-function-1} \eea
Here the positive integer $M$ has to satisfy the following constraint in order to match the the asymptotic behavior  (\ref{Eigen-Asymp})
of $\L(u)$, namely,
\bea
&&\{e^{\theta_+-\theta_-}F^{(+)}F^{(-)}+e^{-(\theta_+-\theta_-)}D^{(+)}D^{(-)}\}-(-1)^{N}e^{-(\alpha_++\beta_++\alpha_-+\beta_-)}G^{(-)}H^{(+)}
e^{-(2N+2M+1)\eta}\no\\[6pt]
&&~~~~~~~~~~~~~-(-1)^{N}e^{(\alpha_++\beta_++\alpha_-+\beta_-)}G^{(+)}H^{(-)}e^{(2N+2M+1)\eta}=0.\label{m-constraint}
\eea
Moreover, the $M$ parameters $\{\l_j|j=1,\cdots,M\}$ need to satisfy the associated BAEs
\bea
\frac{a(\lambda_j)}{d(\lambda_j)}=-\frac{\bar Q(\l_j+\eta)}{\bar Q(\l_j-\eta)},\quad j=1,\cdots,M.\label{BAE-4}
\eea

It is easy to check that the relation (\ref{Constraint-2}) give rise to $F^{(p(2N+4))}=F^{(-p(2N+4))}=0$. Together with (\ref{constraint-9}) and the constrained
case in (\ref{Constraint-2}), we get that the function $F(u)$ indeed vanishes, namely, $ F(u)=0$. Substituting (\ref{T-Q relation-2}) into (\ref{Eigenvlue-2}), we have
\bea
\L^{(\frac{p}{2})}(u)&=&\bar{{\bf {\cal{A}}}}(u)+ \bar{{\bf {\cal{D}}}}(u)
       +\d(u-(\frac{p-1}{2})\eta)\L^{(\frac{p-2}{2})}(u)\no\\
&=&{\bf {\cal{\tilde{A}}}}(u)+ {\bf {\cal{\tilde{D}}}}(u)
       +\d(u-(\frac{p-1}{2})\eta)\L^{(\frac{p-2}{2})}(u).\label{B-2}
\eea
Similar with the periodic case, we can also prove that the reduced $T-Q$ relation (\ref{T-Q relation-2}) satisfies the functional relations (\ref{Eigen-Asymp})-(\ref{special-values2}) of the transfer matrix and the truncation identity of the fused transfer matrices (\ref{eigenvalue-fused})
when  the inhomogeneity parameters and the boundary parameters
$\{d^{(\pm)}_n,~f^{(\pm)}_n,~g^{(\pm)}_n,~h^{(\pm)}_n,~\alpha_{\pm},~\beta_{\pm},~\theta_{\pm} |n=1,\cdots,N\}$  satisfy the constraints (\ref{Constraint-1}), (\ref{Constraint-2}), (\ref{constraint-9}) and (\ref{m-constraint}).

\section{Conclusion}
In this paper, we have studied the most general cyclic representation of the quantum $\tau_2$-model (also known as Baxter-Bazhanov-Stroganov (BBS) model) with
generic integrable boundary conditions via the ODBA method \cite{yu15}.
Based on the truncation identity (\ref{transfer-fused}) of the fused transfer matrices obtained from the fusion technique, we construct the corresponding inhomogeneous $T-Q$ relation (\ref{T-Q-relation}) and the associated BAEs (\ref{BAE-1}) for the eigenvalue of the fundamental transfer matrix $t(u)$ .

It is  remarked that if the generic inhomogeneity parameters
$\{d^{(\pm)}_n, f^{(\pm)}_n, g^{(\pm)}_n, h^{(\pm)}_n $ $|n=1, \cdots, N\}$
(only obey the constraint (\ref{Constraint-1}) which ensures the integrability of the model) and the boundary parameters take the generic values,
the inhomogeneous term (i.e., the third term) in the $T-Q$ relation (\ref{T-Q-relation}) {\it does not}  vanish, as long as one
requires a polynomial $Q$-function. However, if these
inhomogeneity parameters and the boundary parameters satisfy the extra constraints
(\ref{Constraint-2}), (\ref{constraint-9}) and (\ref{m-constraint}), the resulting inhomogeneous
$T-Q$ relation (\ref{T-Q-relation}) reduces to the conventional one (\ref{T-Q
relation-2}).

\textit{Note added:} After this paper was completed we became aware of the recent results reported in \cite{Maillet16}. The authors use the
Sklyanin's  separation of variables (SoV) method \cite{Skl85,Skl95}   to study
the spectral problem for the open $\tau_2$-model with some constrains on inhomogeneous parameters and also on the boundary parameters.

\section*{Acknowledgments}

We would like to thank Prof. Y. Wang for his valuable discussions
and continuous encouragement. The financial supports from the National Natural Science Foundation
of China (Grant Nos. 11375141, 11374334, 11347025, 11434013, 11425522, and
11547045), BCMIIS, 2016YFA0300603, and the Strategic Priority
Research Program of the Chinese Academy of Sciences are gratefully
acknowledged. X. Xu also acknowledges the support by the NWU graduate student
innovation fund No. YZZ14102.

\section*{A~~Specific cases of the fused $K$-matrices}
\setcounter{equation}{0}
\renewcommand{\theequation}{A.\arabic{equation}}
In this appendix we present the explicit expressions of the matrix elements of the fused $K$-matrices
$K^{\pm(\frac{p}{2})}(u)$ given in (\ref{tk-fused}) and (\ref{kp-function}) for the $p=3$ case as an example. In this case,
the corresponding  matrix elements are
\bea
\mathcal{K}_{11}^{-(\frac{3}{2})}(u)&=&(\frac{1}{2})^{2}\Big\{(\sinh\alpha_{-}\cosh\beta_-)^3\cosh(3u)+3(\sinh\alpha_{-}\cosh\beta_{-})^2\cosh\alpha_{-}\sinh\beta_{-}\sinh(3u)\no\\[6pt]
&+&3\sinh\alpha_{-}\cosh\beta_-(\cosh\alpha_{-}\sinh\beta_{-})^2\cosh(3u)+(\cosh\alpha_{-}\sinh\beta_{-})^3\sinh(3u)\no\\[6pt]
&+&3\sinh\alpha_{-}\cosh\beta_{-}\cosh(3u)+3\cosh\alpha_{-}\sinh\beta_{-}\sinh(3u)\Big\},
\eea
\bea
\mathcal{K}_{22}^{-(\frac{3}{2})}(u)&=&(\frac{1}{2})^{2}\Big\{(\sinh\alpha_{-}\cosh\beta_-)^3\cosh(3u)-3(\sinh\alpha_{-}\cosh\beta_{-})^2\cosh\alpha_{-}\sinh\beta_{-}\sinh(3u)\no\\[6pt]
&+&3\sinh\alpha_{-}\cosh\beta_-(\cosh\alpha_{-}\sinh\beta_{-})^2\cosh(3u)-(\cosh\alpha_{-}\sinh\beta_{-})^3\sinh(3u)\no\\[6pt]
&+&3\sinh\alpha_{-}\cosh\beta_{-}\cosh(3u)-3\cosh\alpha_{-}\sinh\beta_{-}\sinh(3u)\Big\},\eea
\bea
\mathcal{K}_{12}^{-(\frac{3}{2})}(u)&=&(\frac{1}{2})^{2}e^{3\theta_-}\sinh(6u),~~~~~~~\mathcal{K}_{21}^{-(\frac{3}{2})}(u)=(\frac{1}{2})^{2}e^{-3\theta_-}\sinh(6u),\\[6pt]
K_{33}^{-(\frac{3}{2})}(u)&=&\frac{{\mathrm{Det}}_{q}\{K^{-}(u-\eta)\}}{\sinh(2u-\eta)}\sigma^{z}K^{-}(u)\sigma^{z},
\eea
and
\bea
\mathcal{K}_{11}^{+(\frac{3}{2})}(u)&=&(\frac{1}{2})^{2}\Big\{-(\sinh\alpha_{+}\cosh\beta_+)^3\cosh(3u)+3(\sinh\alpha_{+}\cosh\beta_{+})^2\cosh\alpha_{+}\sinh\beta_{+}\sinh(3u)\no\\[6pt]
&-&3\sinh\alpha_{+}\cosh\beta_+(\cosh\alpha_{+}\sinh\beta_{+})^2\cosh(3u)+(\cosh\alpha_{+}\sinh\beta_{+})^3\sinh(3u)\no\\[6pt]
&-&3\sinh\alpha_{+}\cosh\beta_{+}\cosh(3u)+3\cosh\alpha_{+}\sinh\beta_{+}\sinh(3u)\Big\},\eea
\bea
\mathcal{K}_{22}^{+(\frac{3}{2})}(u)&=&-(\frac{1}{2})^{2}\Big\{(\sinh\alpha_{+}\cosh\beta_+)^3\cosh(3u)+3(\sinh\alpha_{+}\cosh\beta_{+})^2\cosh\alpha_{+}\sinh\beta_{+}\sinh(3u)\no\\[6pt]
&+&3\sinh\alpha_{+}\cosh\beta_+(\cosh\alpha_{+}\sinh\beta_{+})^2\cosh(3u)+(\cosh\alpha_{+}\sinh\beta_{+})^3\sinh(3u)\no\\[6pt]
&+&3\sinh\alpha_{+}\cosh\beta_{+}\cosh(3u)+3\cosh\alpha_{+}\sinh\beta_{+}\sinh(3u)\Big\},
\eea
\bea
\mathcal{K}_{12}^{+(\frac{3}{2})}(u)&=&-(\frac{1}{2})^{2}e^{3\theta_+}\sinh(6u),~~~~~~~\mathcal{K}_{21}^{+(\frac{3}{2})}(u)=-(\frac{1}{2})^{2}e^{-3\theta_+}\sinh(6u),\\[6pt]
K_{33}^{+(\frac{3}{2})}(u)&=&-\frac{{\mathrm{Det}}_{q}\{K^{+}(u-\eta)\}}{\sinh2u}\sigma^{z}K^{+}(u)\sigma^{z}.
\eea
\section*{B~~Explicit expression of the average value functions}
\setcounter{equation}{0}
\renewcommand{\theequation}{B.\arabic{equation}}
In this appendix we discuss certain properties of the average values of the matrix elements of the monodromy matrices $T(u)$ and $\hat{T}(u)$ given by (\ref{Average-1}) and (\ref{Average-2}) respectively. Here we present  the explicit expressions of these average value functions ${\bf {\cal{A}}}(u),{\bf {\cal{B}}}(u),{\bf {\cal{C}}}(u), {\bf {\cal{D}}}(u)$ and ${\bf {\cal{\hat{A}}}}(u), {\bf {\cal{\hat{B}}}}(u), {\bf {\cal{\hat{C}}}}(u), {\bf {\cal{\hat{D}}}}(u)$ for some small sites cases (namely, $N=1,2$).
For $N=1$, they are given by
\bea
{\bf {\cal{A}}}(u)&=&e^{pu}\{d_1^{(+)}\}^{p}+e^{-pu}\{d_1^{(-)}\}^{p},\\[6pt]
{\bf {\cal{D}}}(u)&=&e^{pu}\{f_1^{(+)}\}^{p}+e^{-pu}\{f_1^{(-)}\}^{p},\\[6pt]
{\bf {\cal{B}}}(u)&=&\{g_1^{(+)}\}^{p}+\{g_1^{(-)}\}^{p},~~~~~{\bf {\cal{C}}}(u)=\{h_1^{(+)}\}^{p}+\{h_1^{(-)}\}^{p},
\eea
\bea
{\bf {\cal{\hat{A}}}}(u)&=&e^{pu}\{f_1^{(-)}\}^{p}+e^{-pu}\{f_1^{(+)}\}^{p},\\[6pt]
{\bf {\cal{\hat{D}}}}(u)&=&e^{pu}\{d_1^{(-)}\}^{p}+e^{-pu}\{d_1^{(+)}\}^{p},\\[6pt]
{\bf {\cal{\hat{B}}}}(u)&=&-\{g_1^{(+)}\}^{p}-\{g_1^{(-)}\}^{p},~~~~~{\bf {\cal{\hat{C}}}}(u)=-\{h_1^{(+)}\}^{p}-\{h_1^{(-)}\}^{p}.
\eea
For $N=2$, they are
\bea
{\bf {\cal{A}}}(u)&=&e^{2pu}\{d_1^{(+)}d_2^{(+)}\}^{p}+e^{-2pu}\{d_1^{(-)}d_2^{(-)}\}^{p}+\{d_1^{(-)}d_2^{(+)}\}^{p}+\{d_1^{(+)}d_2^{(-)}\}^{p}\no\\[6pt]
&+&\Big(\{g_2^{(+)}\}^{p}+\{g_2^{(-)}\}^{p}\Big)\Big(\{h_1^{(+)}\}^{p}+\{h_1^{(-)}\}^{p}\Big),\\[6pt]
{\bf {\cal{D}}}(u)&=&e^{2pu}\{f_1^{(+)}f_2^{(+)}\}^{p}+e^{-2pu}\{f_1^{(-)}f_2^{(-)}\}^{p}+\{f_1^{(-)}f_2^{(+)}\}^{p}+\{f_1^{(+)}f_2^{(-)}\}^{p}\no\\[6pt]
&+&\Big(\{g_1^{(+)}\}^{p}+\{g_1^{(-)}\}^{p}\Big)\Big(\{h_2^{(+)}\}^{p}+\{h_2^{(-)}\}^{p}\Big),\\[6pt]
{\bf {\cal{B}}}(u)&=&e^{pu}\Big\{(\{g_1^{(+)}\}^{p}+\{g_1^{(-)}\}^{p})\{d_2^{(+)}\}^{p}+(\{g_2^{(+)}\}^{p}+\{g_2^{(-)}\}^{p})\{f_1^{(+)}\}^{p}\Big\}\no\\[6pt]
&+&e^{-pu}\Big\{(\{g_1^{(+)}\}^{p}+\{g_1^{(-)}\}^{p})\{d_2^{(-)}\}^{p}+(\{g_2^{(+)}\}^{p}+\{g_2^{(-)}\}^{p})\{f_1^{(-)}\}^{p}\Big\},\\[6pt]
{\bf {\cal{C}}}(u)&=&e^{pu}\Big\{(\{h_1^{(+)}\}^{p}+\{h_1^{(-)}\}^{p})\{f_2^{(+)}\}^{p}+(\{h_2^{(+)}\}^{p}+\{h_2^{(-)}\}^{p})\{d_1^{(+)}\}^{p}\Big\}\no\\[6pt]
&+&e^{-pu}\Big\{(\{h_1^{(+)}\}^{p}+\{h_1^{(-)}\}^{p})\{f_2^{(-)}\}^{p}+(\{h_2^{(+)}\}^{p}+\{h_2^{(-)}\}^{p})\{d_1^{(-)}\}^{p}\Big\},\\[6pt]
{\bf {\cal{\hat{A}}}}(u)&=&e^{2pu}\{f_1^{(-)}f_2^{(-)}\}^{p}+e^{-2pu}\{f_1^{(+)}f_2^{(+)}\}^{p}+\{f_1^{(-)}f_2^{(+)}\}^{p}+\{f_1^{(+)}f_2^{(-)}\}^{p}\no\\[6pt]
&+&\Big(\{g_1^{(+)}\}^{p}+\{g_1^{(-)}\}^{p}\Big)\Big(\{h_2^{(+)}\}^{p}+\{h_2^{(-)}\}^{p}\Big),\\[6pt]
{\bf {\cal{\hat{D}}}}(u)&=&e^{2pu}\{d_1^{(-)}d_2^{(-)}\}^{p}+e^{-2pu}\{d_1^{(+)}d_2^{(+)}\}^{p}+\{d_1^{(-)}d_2^{(+)}\}^{p}+\{d_1^{(+)}d_2^{(-)}\}^{p}\no\\[6pt]
&+&\Big(\{g_2^{(+)}\}^{p}+\{g_2^{(-)}\}^{p}\Big)\Big(\{h_1^{(+)}\}^{p}+\{h_1^{(-)}\}^{p}\Big),\\[6pt]
{\bf {\cal{\hat{B}}}}(u)&=&-e^{pu}\Big\{(\{g_1^{(+)}\}^{p}+\{g_1^{(-)}\}^{p})\{d_2^{(-)}\}^{p}+(\{g_2^{(+)}\}^{p}+\{g_2^{(-)}\}^{p})\{f_1^{(-)}\}^{p}\Big\}\no\\[6pt]
&-&e^{-pu}\Big\{(\{g_1^{(+)}\}^{p}+\{g_1^{(-)}\}^{p})\{d_2^{(+)}\}^{p}+(\{g_2^{(+)}\}^{p}+\{g_2^{(-)}\}^{p})\{f_1^{(+)}\}^{p}\Big\},\\[6pt]
{\bf {\cal{\hat{C}}}}(u)&=&-e^{pu}\Big\{(\{h_1^{(+)}\}^{p}+\{h_1^{(-)}\}^{p})\{f_2^{(-)}\}^{p}+(\{h_2^{(+)}\}^{p}+\{h_2^{(-)}\}^{p})\{d_1^{(-)}\}^{p}\Big\}\no\\[6pt]
&-&e^{-pu}\Big\{(\{h_1^{(+)}\}^{p}+\{h_1^{(-)}\}^{p})\{f_2^{(+)}\}^{p}+(\{h_2^{(+)}\}^{p}+\{h_2^{(-)}\}^{p})\{d_1^{(+)}\}^{p}\Big\}.
\eea
The  results show that the average value functions become tedious  when the number of the lattice sites $N$ goes large.
However, they can be worked out for an arbitrary $N$  when the inhomogeneous parameters $\{g^{(+)}_n,~g^{(-)}_n,~h^{(+)}_n,~h^{(-)}_n\}$ associated with each site $n$ satisfy some constraints.
One case is the parameters $\{g^{(+)}_n,~g^{(-)}_n,~h^{(+)}_n,~h^{(-)}_n|n=1,\cdots,N\}$ obey the chiral Potts constraints \cite{Tar92}
\bea
\frac{\{g_{n}^{(+)}\}^{p}+\{g_{n}^{(-)}\}^{p}}{\{h_{n}^{(+)}\}^{p}+\{h_{n}^{(-)}\}^{p}}=\lambda,
\eea
where $\lambda$ is arbitrary constant. The other case is that the parameters $d^{(+)}_n$, $d^{(-)}_n$, $g^{(+)}_n$, $g^{(-)}_n$, $h^{(+)}_n$, $h^{(-)}_n$, $f^{(+)}_n$ and $f^{(-)}_n$ are independent of the site (i.e., $n$), which corresponds to the homogeneous model.  In the both cases, the average values of $L$-operators $\{{\cal{L}}_n(u)|n=1,2,\cdots,N\}$ given in (\ref{Average-3}) (or  $\{\hat{{\cal{L}}}_n(u)|n=1,2,\cdots,N\}$ given in (\ref{Average-4})) with different sites
commute with each other. Thus one can diagonalize them simultaneously. Then the average values of the elements of monodromy matrices (\ref{Average-1}) and (\ref{Average-2}) can be easy to obtain.


\end{document}